 \documentclass[aps,prl,groupedaddress,nofootinbib,reprint]{revtex4-1}
\usepackage{times}
\usepackage{bm}
\usepackage{graphicx}
\usepackage{color}
\usepackage{ulem}

\DeclareMathAlphabet{\mathscr}{OT1}{pzc}{m}{it}

\newcommand{\expect}[1]{{\left\langle #1 \right\rangle}}

\newcommand{\dens}[1]{{\rm [#1]}}

\def \expect#1{{\left \langle #1 \right\rangle}}

\newcommand{\be}{\begin{eqnarray}}
\newcommand{\ee}{\end{eqnarray}}

\def\Thad#1{#1}

\def\lr#1{{\left(#1\right)}}

\begin{document}

 \title{Suppression of Spin-Exchange Relaxation Using Pulsed Parametric Resonance}

\def\Wisc{Department of Physics, University of Wisconsin-Madison, Madison, WI 53706, USA}
\def\Julich{Juelich Centre for Neutron Science, Garching 85747, Germany
}
\author{A. Korver}
\author{ R. Wyllie}\altaffiliation[Current address: ]{National Institute of Standards and Technology, Gaithersburg, MD.}
\author{ B. Lancor}
\author{T. G. Walker}
\affiliation{\Wisc}

\date{\today}

\begin{abstract} 

\Thad{We demonstrate that spin-exchange dephasing of Larmor precession at near-earth-scale fields  is effectively eliminated by dressing the alkali-metal atom spins in a  sequence of AC-coupled 2$\pi$ pulses, repeated at the Larmor precession frequency.  The 
contribution of spin-exchange collisions to the spectroscopic line width is reduced by a factor of the duty cycle of the pulses.  We experimentally demonstrate  resonant transverse pumping  in magnetic fields as high as 0.1 Gauss, present experimental measurements of the suppressed spin-exchange relaxation, and show  enhanced magnetometer response relative to a light-narrowed scalar magnetometer.}

\end{abstract}

\maketitle


\Thad{Many precision measurement devices are based upon optically pumped  high density alkali-metal vapors \cite{HJW} confined in glass cells. The long coherence times (10 msec--60 sec \cite{Balabas2010}) of these spin-polarized atoms, combined with sensitive optical spin-detection, enable compact  high resolution atomic clocks and magnetometers. The frequency resolution attainable in alkali-metal spectroscopy is normally limited by spin-exchange collisions that dominate at high alkali-metal atom densities. Spin-exchange decoherence can be suppressed by operating near zero absolute magnetic field in the spin-exchange relaxation free (SERF) regime\cite{Happer1973,Happer1977}, where alkali-metal densities can be increased by orders of magnitude without degradation of the spin coherence times. Rapid developments have followed the first demonstration of a SERF magnetometer \cite{Kominis2003}, including unprecedented magnetic sensitivity \cite{Dang2010}, biomagnetic sensing \cite{Xia06,Bison2009,Johnson2010,Knappe2010,Sander2012,Wyllie2012a,Wyllie2012}  and chip-scale miniaturization \cite{Sander2012}. At higher earth-scale fields, spin-exchange decoherence can be limited in the light narrowing regime, where at high polarization angular momentum conservation prevents relaxation from spin-exchange collisions. The transverse relaxation rate for $^{87}$Rb, $\Gamma_2 \approx\Gamma_{SE}(1-P)/5$  becomes small, but remains an important limiting factor in clocks and magnetometers \cite{Jau04c,  Smullin2009}.}

\Thad{In this Letter, we describe a modulation technique to suppress spin-exchange decoherence at large fields where the traditional SERF mechanism does not apply. By dressing the atomic spins in a sequence of very short AC-coupled 2$\pi$ magnetic field pulses, the spin-exchange decoherence can be decreased by nearly the duty cycle of the pulse, $\Gamma_2\approx 0.3d(1-P) \Gamma_{SE}$. We perform transverse optical pumping in the dressed field, a generalization of transverse parametric resonance with sinusoidal dressing \cite{Slocum1965,DupontRoc1969}. Using a pulse duty cycle of 5\%, we demonstrate an order of magnitude decrease in the spin-exchange decoherence rate.  We also  show a $>30\times$ enhancement in magnetic response when compared with a traditional light-narrowed scalar magnetometer operating in the same ~0.1G field.}

\begin{figure}[htb]
\includegraphics[width=3.3 in]{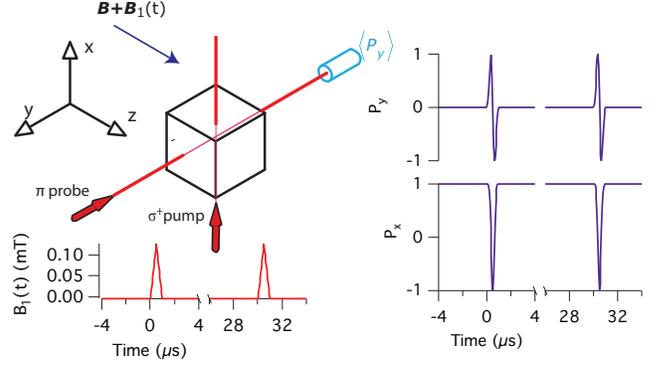}
\caption{\Thad{Transverse optical pumping with pulsed parametric resonance.  A modulating field $\bm B_1$ consists of a series of short $2\pi$ pulses separated in time by the DC Larmor precession period $1/\gamma B$.}  The atoms become polarized along the $\hat x$-direction, and the \Thad{time-averaged} $\hat y$ polarization is sensitive to any differences between the pulse repetition frequency and the DC Larmor frequency. }\label{fig:transmit}
\end{figure}

  Figure~\ref{fig:transmit} is a schematic of \Thad{transverse optical pumping using}  pulsed parametric resonance (PPR)
.  
We  optically pump alkali-metal atoms of gyromagnetic ratio $\gamma$ using D1 $\sigma^+$ light propagating along the $\hat x$-axis, perpendicular to a  DC magnetic field $\bm B=\Omega_z/(2\pi \gamma)\hat z$.
We superpose a \Thad{} modulating field $\bm B_1(t)=\Omega_1(t)/(2\pi \gamma)\hat z$ whose repetition frequency is $f_1=\omega_1/2\pi\approx\Omega_z/2\pi$ and whose time-average is zero.  Neglecting nuclear spin for the moment, the 
  components of the spin polarization $\bf P$ that are transverse to the parametric field, described by $P_+=P_x+iP_y$, obey  the Bloch equation 
\be
{dP_+\over dt}=i(\Omega_{z}+\Omega_1(t))P_+-\Gamma_2 P_++R
\ee
where $\Gamma_2$ is the transverse spin-relaxation rate which includes dephasing processes such as the optical pumping at rate $R$, collisional spin-relaxation, \Thad{diffusion}, and spin-exchange collisions.
Transforming  to a coordinate system that is instantaneously rotated about the z-axis by an angle $\phi_1(t)=\int\Omega_1(t)dt$, {\it i.e.} letting  $P_+=A_+\exp[i\phi_1]$, we get
\be
{dA_+\over dt}=(i\Omega_z-\Gamma_2)A_++Re^{-i\phi_1}
\ee
Expanding
 $\{A_+,\exp[i\phi_1]\}=\sum_p\{A_{+p},j_p\}\exp[i p \omega_1 t]$
gives an equation for the $p$th Fourier coefficient
\be
{dA_{+p}\over dt}=i(\Omega_z-p\omega_1+i\Gamma_2)A_{+p}+Rj^*_{- p}
\ee
Assuming $\Omega_z\gg \Gamma_2$, the $p=1$ Fourier component is resonantly enhanced over the others, giving  the steady-state solution 
\be
P_+\approx{Rj^*_{- 1}\over \Gamma_2-i( \Omega_z-\omega_1)}e^{i\omega_1t+i\phi_1}.
\ee
In particular, the DC spin polarization is
\be
\bar P_{+}={R|j_{- 1}|^2\over \Gamma_2-i( \Omega_z-\omega_1)}
\ee
For a sinusoidal $\Omega_1(t)$,  the $j_p=J_p(|\Omega_1|/\omega_1)$ are Bessel functions, and the maximum DC polarization that can be attained is $P_+=J_1^2(1.84)\approx 0.34$.  Near this maximum, the polarization depends weakly on the magnitude of the modulating field $\Omega_1$ but is very sensitive to the frequency being on resonance.  Thus the transverse spin-polarization, in particular $\bar P_y$, is a sensitive magnetometer for $z$ magnetic fields.

The transverse spin-polarization can be increased by using a shaped magnetic field to increase the size of the Fourier coefficient $j_{-1}$.  The extreme case is a series of ac-coupled $\delta$-functions, {\it i.e.} $B_1=b_1[-1+{\rm comb}(f_1 t)]$.  One gets
\be
j_{-1}=-\text{sinc}\left[\pi(1-{\gamma b_1/ f_1})\right]
\ee
This equals one when $\gamma b_1=f_1$,  corresponding to the atoms experiencing a repetitive sequence of  $2\pi$ rotations about the $\hat z$-axis.  When the repetition frequency also meets the  resonance condition $f_1=\gamma B_z$, the atoms can be fully polarized {\it transverse} to the static magnetic field.  On a time-averaged basis, \Thad{the atoms have a fixed orientation with respect to the static field} 
, as if they effectively have zero magnetic moment.
This is because when $b_1=B_z$  the total magnetic field between the pulses is cancelled.  Thus the following picture of the optical pumping and precession emerges:  when the sequence of pulses meets the conditions $b_1=B_z=f_1/\gamma$, the atoms experience zero magnetic field between the pulses so they do not precess, allowing effectively  zero-field optical pumping.  When each magnetic field pulse arrives, the atoms precess by 2$\pi$ about the $\hat z$-axis.  

\Thad{More important than the factor $\sim3$ increase in the maximum attainable polarization for transverse pumping is the elimination of spin-exchange relaxation between the pulses.  Just as in a SERF magnetometer, spin-exchange collisions at zero magnetic field do not contribute to  relaxation.  }
Only collisions that occur during the short $2\pi$ pulses contribute to the transverse relaxation rates.  
 \Thad{The reduction in transverse relaxation thus greatly reduces the pumping rate required to attain full transverse polarization. } At high spin-exchange rates a PPR magnetometer, with its greater polarization and \Thad{narrower} line width, will have a superior response as compared to other types of transversely pumped  magnetometers 

\Thad{For pulses of finite duration, the maximum attainable DC polarization is less than one.}
Retaining the definition of $b_1$ as the negative of the AC magnetic field between the pulses, and assuming square pulses of duty cycle $d$, the maximum polarization  is reduced to $|j_{-1}|^2=(1-d)^2$, arising from the rotation of the spins during the pulses. 

Inclusion of nuclear spin angular momentum $\bm I$ adds additional complications. As long as quadratic Zeeman effects can be neglected, the nuclear spin inertia simply reduces the gyromagnetic ratio by a factor of $(2I+1)$ as compared to an electron.  The magnetic field pulse area must be increased by the same factor to achieve the 2$\pi$ rotations in the same amount of time.

We now consider the effects of spin-exchange collisions during the pulses. Spin-exchange collisions conserve the total angular momenta but can transfer populations and coherences between the different hyperfine levels that have equal and opposite magnetic moments.  In non-zero magnetic fields, the opposing precession directions imply that spin-exchange collisions rapidly dephase the Larmor precession, unless spin-exchange collisions occur at such a high rate that the atoms precess with a smaller averaged Larmor frequency\cite{Happer1977}.  In low magnetic fields, such as in a SERF magnetometer or between the pulses of the PPR magnetometer, the spin-exchange induced transfer between the two hyperfine levels is of little consequence.

\begin{figure}[thb]
\includegraphics[width=3.5 in]{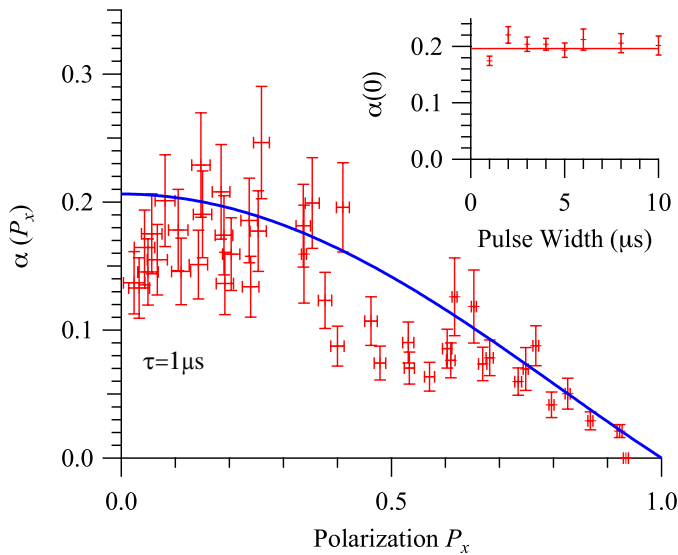}
\caption{Fraction of angular momentum lost when a spin-exchange collision occurs during a 2$\pi$ magnetic field pulse as a function of $x-$polarization or pulse width $\tau$ (inset).  The solid line is the model prediction; the data are results deduced from polarization and relaxation measurements.} \label{fig:alpha}
\end{figure}

In contrast, spin-exchange collisions during the 2$\pi$ pulses are important contributors to dephasing.    In a (primed) reference frame rotating at the hyperfine Zeeman precession frequency,  $\expect{F_x'}$ is conserved but spin-exchange acts to equalize the populations of states $|I+1/2,m'\rangle$ and $|I-1/2,m'\rangle$.  If a spin-exchange collision occurs after precession by angle $\phi$ during the pulse, and moves a particular atom from one hyperfine level to the other, the  sign of its magnetic moment flips.  It will therefore subsequently precess an angle $2\pi-\phi$ in the wrong direction, with a consequent rotation error of $2\phi$ at the end of the pulse.  Averaging over the  distribution of angles at which the spin-exchange occurs, and assuming that multiple spin-exchange collisions during the pulses can be neglected, an analysis using the techniques of Ref. \cite{HJW} predicts  an average fractional angular momentum loss per spin-exchange collision of
$
{\Delta  F_x/  F_x}=\alpha
$
where $\alpha\approx\alpha'(1-P_x)$ at high polarization ($\alpha'\approx0.3$) , and increases to $\alpha=0.21$ at low polarizations,  as shown in Fig.~\ref{fig:alpha}.  Here and elsewhere our  results are specific to $^{87}$Rb with $I=3/2$. Averaged over many pulses, we therefore expect spin-exchange collisions during the pulses to produce an effective spin-relaxation rate of
\be
{d\bm F\over dt}=-\alpha  d \Gamma_{SE}\bm F=-\Gamma_p \bm F
\ee
Relaxation due to spin-exchange collisions is not only suppressed at high polarization, similar to the light-narrowing phenomenon, but is also reduced by only having a fraction $d$ of spin-exchange collisions contribute to the relaxation.
Connecting back to the language of the Bloch magnetic resonance picture, the  PPR magnetometer has an effective relaxation rate from spin-exchange collisions of  $\Gamma_2=d\alpha\Gamma_{SE}$. 

We turn now to the experimental demonstration of transverse optical pumping with pulsed-field parametric modulation.  We perform D1 optical pumping of a 87-Rb cell containing 200 Torr of N$_2$ buffer gas, with the pumping laser propagation direction perpendicular to a DC magnetic field $B_0=0.04$ G in the $\hat z$-direction.  This field is generated by magnetic field coils inside a magnetic shield.   A second set of coils produce the pulsed magnetic field parallel to $\bm B_0$.  These 4 $\mu$H coils, in series  with a 22 $\mu$F capacitor, are driven by a pulsed MOSFET circuit with a Q-spoiling 200 $\Omega$ resistor to produce a series of AC-coupled triangular magnetic field pulses of  $\tau=1.0$ $\mu$s duration.  To reduce the effects of eddy currents in the magnetic shield, the ac field coil set is designed to have have zero magnetic moment.  Also, series inductors in the DC field coil circuit substantially reduce the effects of mutual inductance on the resulting field pulses. 

\begin{figure}[htb]
\includegraphics[width=3.5 in]{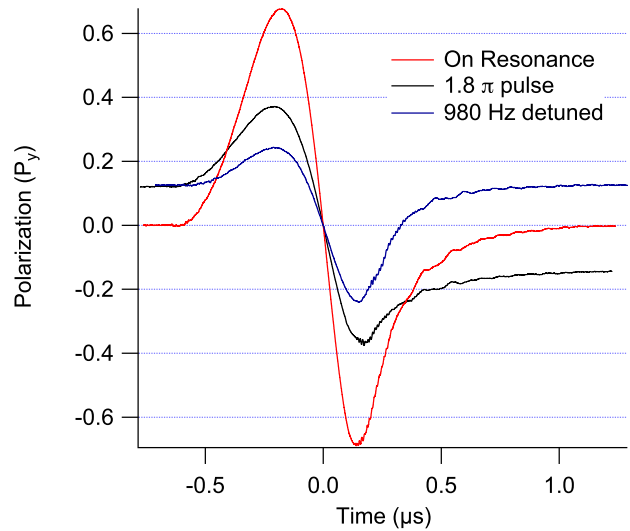}
\caption{Observed y-polarization waveforms observed with pulsed parametric modulation.  Pulses of the proper repetition rate and area cause rapid 2$\pi$ precession of the spins, returning them to he pulses, with no precession between pulses.  If the pulse area is incorrect, there is a residual magnetic field between the pulses that causes inter-pulse precession.  If the repetition frequency is off-resonance, the atomic polarization is rotated away from the $\hat x$ direction.}\label{fig:pulse}
\end{figure}

The optical pumping beam is  tuned several line widths off-resonance to minimize optical thickness effects and produce a relatively uniform pumping rate throughout the cell. A second off-resonant {(0.2 nm)} probe laser senses the $\hat y$-component of the electron polarization $P_y$ using Faraday rotation.  When the pulsed magnetic field is tuned to be resonant and have the correct area, we observe $P_y$ to have the expected shape shown in Fig.~\ref{fig:pulse}.  This figure also shows waveforms observed when the pulsed field frequency is off-resonance, and when the pulse area is non-optimum.  Again, the qualitative features expected are observed.

As a quantitative study, in particular to test the predicted contributions of spin-exchange collisions to the transverse relaxation rate, we measure the polarization obtained as a function of pulse width, pulse frequency, pumping rate, and density.  To avoid complications of keeping the pulse frequency on resonance, we set $B_0=0$ and remove the ac-coupling capacitor so our time-dependent magnetic field becomes simply a series of 2$\pi$ pulses.  We shim the transverse magnetic fields to nearly zero as well. Assuming spin-temperature equilibrium, the optical pumping process can be represented as a statement of conservation of angular momentum:  
\be
{dF_x\over dt}={R|j_{-1}|^2\over 2}-\Gamma'{P_x\over 2}-\Gamma_pF_x
\label{spintempevol}
\ee
where $\Gamma'=\Gamma_0+R$ includes collisional, optical pumping, and other relaxation rates.
In spin-temperature, the spin-polarization $P_x$ (deduced from the amplitude of the signals in Fig.~\ref{fig:pulse}) and the total angular momentum $F_x=qP_x/2$ are related by a slowing-down factor $q$ ($4<q<6$ for $^{87}$Rb) where $1/q$ is the fraction of the total angular momentum stored in the electron spin.  This can be written in the Bloch equation form with the definition $\Gamma_2=\Gamma'/q+\alpha d \Gamma_{SE}$. Solving Eq. (\ref{spintempevol}) in  steady state, the pulse-induced relaxation can be deduced from the polarization $P_x(f)$ at pulse frequency $f$ as follows:
\be
{\Gamma_p}={\Gamma'\over q(P_x)}\lr{{P_x(0)\over P_x(f)}-1}=\Gamma_{p0}+\alpha(P_x)\tau fk_{SE}\dens{Rb}
\label{Gammap}
\ee
All quantities in this equation (excepting $\alpha$) are either measured or, for $q(P_x)$ and $k_{SE}$, well-known from the properties of spin-temperature or prior experiments \cite{Walter2002}.
We calibrate the polarization measurements using $2\pi$-pulse amplitudes in the limits $f\rightarrow0$, $R\gg \Gamma_0$.  We measure $\Gamma_0$ from relaxation in the dark, and $R$ from the intensity dependence of $P_x(0)$.  We measure the density \dens{Rb} using the absolute magnitude of the Faraday rotation signal.  In addition to spin-exchange loss, we allow for an observed small amount $\Gamma_{p0}$ of non spin-exchange loss presumably due to pulse imperfections.

Figure~\ref{fig:alpha} shows the fractional angular momentum loss deduced  for 1 $\mu$s pulses, for a variety of pumping rates, repetition frequencies (1-30 kHz), and densities ($10^{12}$--$10^{14}$/cm$^3$).  With no free parameters, the results are close to the expected values.  Allowing for an overall scaling of $\alpha(0)$, the collection of data for pulse widths ranging from 1 $\mu$s to 8 $\mu$s gives a mean value $\alpha(0)=0.20\pm0.01$.  In summary, the data/model agreement confirm that the spin-exchange relaxation under PPR conditions is suppressed by a factor of $\alpha d$.

As another demonstration of suppression of spin-exchange relaxation, we now compare a PPR magnetometer to other atomic magnetometers that have been under intense development in the past few years.  To this end, we align the pumping light along the $\hat{z}$-axis, and sense the response to small transverse magnetic fields.  Accounting for residual spin-exchange relaxation,  the on-resonance PPR magnetometer response  is
\be
{\cal M}=\dens{Rb}{dP_y\over d\Omega_x}=\dens{Rb}{R|j_{-1}|^2\over \Gamma'(\Gamma'+q\Gamma_p)}
\ee
Because $\Gamma_p$ decreases with increasing pumping rate due to the decreased spin-exchange loss at high polarizations, the optimum pumping rate $R= \sqrt{\Gamma_0 \alpha'qd\Gamma_{SE}\Thad{+\Gamma_0^2}}$ is higher than for a SERF.  The magnetometer response optimizes at 
\be
\mathcal{M}=
\frac{\dens{Rb}|j_{-1}|^2}{2  \left(\sqrt{{\Gamma_0}q\alpha' d \Gamma_{SE}+{\Gamma_0}^2}+{{\Gamma_0}}\right)}
\ee
For comparison, the optimum SERF response, obtained when $R=\Gamma_0$, gives ${\cal M}=\dens{Rb}/4\Gamma_0$.  At very high spin-exchange rates, $\Gamma_{SE}=2\times 10^5$/s, $\Gamma_0=500$/s, $d=0.05$, the ratio of the \Thad{measured}  {\it zero-field} SERF response to the PPR response at 0.43 $\mu$T is 3.5, in reasonable agreement with the expected 3.0.

\begin{figure}[htb]
\includegraphics[width=3.2 in]{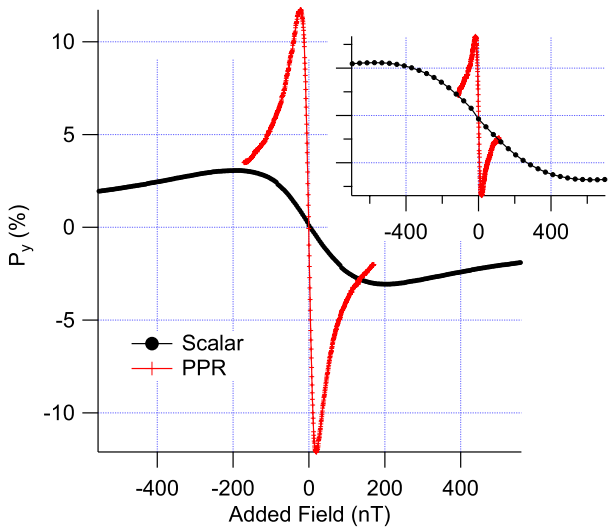}
\caption{Comparison of magnetometer responses for scalar and PPR magnetometers. The main figure shows an optimized experimental comparison at high optical pumping rates, where spin-exchange collisions are partially mitigated for both magnetometers.  \Thad{The magnetic response on resonance is $33\times$ larger for the PPR magnetometer.}The inset, taken at low pumping rates, shows directly the reduced effects of spin-exchange for the PPR case.  }\label{fig:response}
\end{figure}

For operation at magnetic fields $\Omega_z\gg\Gamma_0$ as considered in this paper, a more relevant comparison is to scalar magnetometers.  Light-narrowing suppresses spin-exchange relaxation at high polarizations \cite{Happer1977} in a manner analogous to the $1-P$ behavior of $\alpha$ (Fig.~\ref{fig:alpha}). A detailed analysis of the response of a scalar magnetometer at high densities was presented by Smullin {\it et al.}\cite{Smullin2009}.   
Figure~\ref{fig:response} shows an experimental comparison of the two magnetometers under the same conditions as the SERF comparison discussed above.  The two magnetometers were separately optimized for 
pumping \Thad{rate} and, in the scalar case, optimum rf field amplitude.  The ratio of the two responses for small field deviations is 33. The inset shows \Thad{how the PPR spin-exchange suppression reduces the resonance line width at low polarization}. 



By dressing alkali-metal atoms in a  resonant pulsed magnetic field, we have shown that spin-relaxation due to spin-exchange collisions can be dramatically reduced. The atoms can be optically pumped as if in zero-field, and have magnetic field sensitivities in near-earth-scale fields that approach zero-field spin-exchange-free levels.  
 With appropriate additional modulations, it should be possible to extend these results to optimize AC response at kHz frequencies that may be of particular interest to low-field MRI/NMR detection.  

We acknowledge help and advice from M. Larsen, M. Kauer, M. Ebert, and I. Sulai.  This work was supported by the NSF and  the National Institutes of Health Eunice Kennedy Shriver National Institute of Child Health \& Human Development, R01HD057965.

\vspace*{1.0 in}

\bibliography{/Users/Thad_Walker/Research/thadbibtex/spinexchange}

\end{document}